\documentclass[11pt]{article}

\usepackage{amsmath}
\usepackage{amssymb}
\usepackage{amsthm}
\usepackage[margin=0.75in]{geometry}
\usepackage{graphicx}
\usepackage{subcaption}
\usepackage[round]{natbib}
\usepackage{multicol}
\usepackage{tabularx}

\usepackage{xcolor}
\usepackage{blindtext}
\usepackage{hyperref}

\definecolor{tab_blue}{HTML}{4878D0}
\definecolor{tab_orange}{HTML}{EE854A}
\usepackage[acronym,nowarn]{glossaries}

\usepackage{algorithm}
\usepackage{algpseudocode}
\usepackage{inconsolata}
\usepackage{siunitx}

\graphicspath{{./figures/}}

\newacronym{NN}{NN}{neural network}
\newacronym{GP}{GP}{Gaussian process}

\newacronym{statFEM}{statFEM}{statistical finite element method}
\newacronym{EnKF}{EnKF}{ensemble Kalman filter}
\newacronym{ExKF}{ExKF}{extended Kalman filter}
\newacronym{DA}{DA}{data assimilation}
\newacronym{SDE}{SDE}{stochastic differential equation}
\newacronym{ODE}{ODE}{ordinary differential equation}
\newacronym{PDE}{PDE}{partial differential equation}
\newacronym{FEM}{FEM}{finite element method}

\newacronym{QoI}{QoI}{quantities-of-interest}
\newacronym{SWE}{SWE}{shallow water equations}
\newacronym{RMSE}{RMSE}{root-mean-square error}
\newacronym{RRMSE}{RRMSE}{relative root-mean-square error}
\newacronym{MSE}{MSE}{mean-squared-error}
\newacronym{LR-ExKF}{LR-ExKF}{low-rank extended Kalman filter}

\def\RMSE{{\mathrm{RMSE}}}
\def\BF{{\mathrm{BF}}}

\def\dee{\,{\mathrm{d}}}

\def\cF{{\mathcal F}}
\def\cD{{\mathcal D}}

\def\cM{{\mathcal M}}

\def\bR{{\mathbb R}}

\def\NPDF{{\mathcal N}}

\def\f0{{\mathbf 0}}

\def\vzero{{\mathbf{0}}}

\def\vr{{\mathbf{r}}}

\def\vu{{\mathbf{u}}}

\def\vx{{\mathbf{x}}}
\def\vy{{\mathbf{y}}}

\def\mA{{\mathbf{A}}}
\def\mB{{\mathbf{B}}}

\def\mG{{\mathbf{G}}}
\def\mH{{\mathbf{H}}}
\def\mI{{\mathbf{I}}}
\def\mJ{{\mathbf{J}}}
\def\mK{{\mathbf{K}}}
\def\mL{{\mathbf{L}}}
\def\mM{{\mathbf{M}}}

\def\mR{{\mathbf{R}}}

\def\mU{{\mathbf{U}}}
\def\mV{{\mathbf{V}}}

\def\mLambda{{\mathbf{\Lambda}}}

\def\Dt{{\Delta_t}}
\def\GP{{\mathcal{GP}}}
\def\RMSE{{\mathrm{RMSE}}}

\newcommand{\aprio}{\textit{a priori}}
\newcommand{\apost}{\textit{a posteriori}}

\newcommand{\bm}{\boldsymbol}

\newcommand{\given}{\:\vert\:}

\usepackage{authblk}
\title{Exploring Model Misspecification in Statistical Finite Elements via Shallow Water Equations}
\author[1]{Connor Duffin\footnote{Corresponding author. Email: \texttt{cpd32@cam.ac.uk.}}}
\author[2,3]{Paul Branson}
\author[3]{Matt Rayson}
\author[1,4]{Mark Girolami}
\author[5]{Edward Cripps}
\author[5]{Thomas Stemler}
\affil[1]{Department of Engineering, University of Cambridge, Cambridge, UK}
\affil[2]{CSIRO Environment, Perth, Australia}
\affil[3]{Oceans Graduate School, The University of Western Australia, Perth, Australia}
\affil[4]{The Alan Turing Institute, London, UK}
\affil[5]{Department of Mathematics and Statistics, The University of Western Australia, Perth, Australia}
\makeatletter
\newenvironment{tablehere}
  {\def\@captype{table}}
  {}

\begin{document}

\maketitle

\begin{abstract}
The abundance of observed data in recent years has increased the number of statistical augmentations to complex models across science and engineering.  By augmentation we mean coherent statistical methods that incorporate measurements upon arrival and adjust the model accordingly. However, in this research area methodological developments tend to be central, with important assessments of model fidelity often taking second place.  Recently, the statistical finite element method (statFEM) has been posited as a potential solution to the problem of model misspecification when the data are believed to be generated from an underlying partial differential equation system. Bayes nonlinear filtering permits data driven finite element discretised solutions that are updated to give a posterior distribution which quantifies the uncertainty over model solutions. The statFEM has shown great promise in systems subject to mild misspecification but its ability to handle scenarios of severe model misspecification has not yet been presented. In this paper we fill this gap, studying statFEM in the context of shallow water equations chosen for their oceanographic relevance. By deliberately misspecifying the governing equations, via linearisation, viscosity, and bathymetry, we systematically analyse misspecification through studying how the resultant approximate posterior distribution is affected, under additional regimes of decreasing spatiotemporal observational frequency. Results show that statFEM performs well with reasonable accuracy, as measured by theoretically sound proper scoring rules.
\end{abstract}


\begin{multicols}{2}
\section{Introduction}
\label{sec:intro}

In a crude sense every physical model is
misspecified~\citep{box1979robustness}. Approximations and intentional
omission of processes are necessary in order to build tractable
mathematical representations of reality, however this leads to model
discrepancies when comparisons to observations are
drawn~\citep{oreskes1994verification,judd2004indistinguishable}. Thus
the phenomenon of model misspecification, whereby the data show
inconsistencies with the model employed, is ubiquitous througout engineering
and the physical sciences.

Bayesian statistical approaches, where implementable, provide an
optimal solution to rectify this mismatch with
data~\citep{berger2019statistical}. In such an approach, the posterior
probability distribution over any unknown quantities-of-interest is
estimated. When the quantity-of-interest is the model state, this
estimation is typically the \textit{data assimilation} problem with
relevant posterior distributions being the filtering or smoothing
distributions~\citep{wikle2007bayesian}. In such an aproach, model
uncertainties are typically assumed to be extrusive to the physical
model. Solving the \textit{inverse problem} allows for a similar
estimation however uncertainty in such models are taken inside the
physical model, such as model parameters, initial conditions, or
boundary conditions. See~\cite{stuart2010inverse} for a summary in the
infinite-dimensional setting.

The combination of intrusive model parameter estimation and
extrusive additive model error was formalised as
\textit{Bayesian calibration} in the seminal work
of~\cite{kennedy2001bayesian} (for a review of recent works see
also~\cite{xie2021bayesian}). This additive error was modelled via a
\gls*{GP}, a common and flexible tool which allows for uncertainty
over functions to be modelled in an interpretable
fashion~\citep{williams2006gaussian}.

Adjacent to these works is the recently proposed
\gls*{statFEM}~\citep{girolami2021statistical}; a statistically
coherent Bayesian procedure which updates finite element discretised
\gls*{PDE} solution fields with observed data. Different to previous
works, model errors are intrusive, with \gls*{GP} priors placed on
model components which are potentially unknown, for example
external forcing processes or diffusivity. This uncertainty is then
leveraged to update \gls*{PDE} solutions in an online fashion, to
compute an approximate Gaussian posterior measure using classical
nonlinear filtering algorithms~\citep[see,
e.g.,][]{law2015data}. Previous
work~\citep{duffin2021statistical,duffin2022lowrank} has focused on
applying the \gls*{EnKF} or \gls*{ExKF}, demonstrating the methodology
on canonical systems. Results show that this approach can correct
for model mismatch with sparse observations, allowing the reconstruction
of these phenomena using an interpretable and statistically coherent
physical-statistical model. An interpretation of the methodology is
that it provides a physics-based interpolator which can be applied to
models where assumptions of stationarity may not necessarily
hold. This enables the application of simpler physical models,
correcting for their behaviour with observed data.

However, as yet there has been no systematic analysis of
\gls*{statFEM} under varying degrees of model misspecification. Work
so far has been limited to situations where the posited dynamics
well-approximates the data generating process with either model
parameters or initial conditions having minor perturbations from the
truth. In this work we fill this gap through studying \gls*{statFEM}
in regimes of increasing model misspecification. Using the \gls*{SWE}
as the example system, we deliberately misspecify model
parameters from the known values which are used to generate the data
(in this case, the model viscosity and bathymetry) to see how the
method performs in these various regimes.

We detail a suite of simulation studies to analyse how mismatch
\aprio{} can be corrected for, \apost{}. We also study how linearising
the governing equations and reducing the observation frequency affects
inference. Our results show that
\begin{enumerate}
\item increasing the observational frequency, in both space and time,
  results in reduced model error, with notable improvements as more
  spatial locations are observed
\item misspecifying bathymetry tends to result in less model error
  than viscosity
\item linearising the model may ameliorate some degree of model
  in error if parameters are poorly specified.
\end{enumerate}
We acknowledge that the \gls*{SWE} may not include more highly
nonlinear behaviour that one would expect to see in real-life
settings. However as we investigate joint parameter and
linearisation misspecification a desideratum was such that linear dynamics would approximate the true dynamics.

From a statistical perspective, we are interested in how robust
\gls*{statFEM} is to model misspecification. As such our study follows
the statistical description where our parameters
$\mLambda_{\text{DGP}}$, which generate the data $\vy$, are not the
same as those used to compute the posterior over the model state
$\vu$, $p(\vu \given \vy, \mLambda)$. Our likelihood is thus
misspecified as $p(\vy \given \mLambda) \neq p(\vy \given
\mLambda_{\text{DGP}})$~\citep[see, e.g.,][and the references
therein]{white1982maximum}. We study misspecification in this setting
as, in reality, our models will be misspecified and inference will
never be performed in the so-called ``perfect model
scenario''~\citep{judd2001indistinguishablea}. Parameters and
topography are in reality never known and approximations will need to
be made. Furthermore, linear approximations are often
employed~\citep[see, e.g., ][]{cvitanovic2016chaos} and their use with
\gls*{statFEM} is desirable as the resultant posterior distributions
can be computed exactly (using the Kalman filter) without the need for
linearising the prediction step. Our results are thus relevant for
many contexts in which linear approximate models are
employed. Synthetic data provides the appropriate setting as we can
control the severity of misspecification, without the obfuscation from
additional model approximations involved when modelling experimental
or \textit{in situ} measurements.

Assimilation of data into $1D$ shallow water equations has so far focussed on
bathymetry
inversion~\citep{gessese2011reconstruction,khan2021variational,khan2022variational},
analysis of error covariance
parameterisations~\citep{stewart2013data}, and, the convergence of
schemes with sparse surface height
observations~\citep{kevlahan2019convergence}. Previous work on
\gls*{statFEM}~\citep{girolami2021statistical,duffin2021statistical,duffin2022lowrank}
has demonstrated that under cases of mild misspecification, solutions
to nonlinear and time-dependent \glspl*{PDE} can be corrected for with
data, to give an interpretable posterior distribution. In these
previous works, misspecification was due to either deliberately
incorrect parameters, initial conditions, or missing physics. However
these studies were necessarily focussed on methodological
developments, and did not include comprehensive analyses of
\gls*{statFEM} model misspecification.

This systematic analysis is the focus of this paper. Using the
\gls*{SWE} as the example system, we demonstrate our results using
similar experimental designs as the \gls*{SWE} data assimilation works
detailed above \citep[such
as][]{gessese2011reconstruction,stewart2013data,kevlahan2019convergence}. We
study how misspecification effects the filtering posterior
distribution across a variety of parameter values and observation
patterns, and also provide comparisons between linear approximations
and fully nonlinear models. Different to the previous \gls*{statFEM}
works we study the performance as the degree of model misspecification
is varied from mild to severe; model performance is assessed through
the log-likelihood and the root mean square error scoring
rules~\citep{gneiting2007strictly}.

The paper is structured as follows. In Section~\ref{sec:modelling} we
cover an overview of the \gls*{SWE} model and the \gls*{statFEM}
methodology we employ to condition on data. This includes the
numerical scheme employed and the chosen \gls*{GP} priors over unknown
model components. In Section~\ref{sec:experimental-setup} we outline
the general procedure of the experiments. We detail how the data are
generated, how much noise is added, what the \gls*{GP} hyperparameters
are set to, and for which viscosity and bathymetry parameters the
linear and nonlinear models are run with. In Section~\ref{sec:results}
we detail the results across four subsections. In
Section~\ref{sec:intro-examples} we look at four posterior
distributions, computed for cases of mild misspecification and
spatiotemporal observation frequencies, to provide some intuition for
how the models are performing. In Section~\ref{sec:obs-frequency} we
look at how varying spatiotemporal observation frequency affects the
estimated posterior distribution. Similar analyses of physical
parameter misspecification and linearisation are included in
Sections~\ref{sec:parameter-misspecification}
and~\ref{sec:linearisation}, respectively. The results are discussed
and the paper is concluded in
Section~\ref{sec:conclusion}.
For quick reference the paper structure is given in
Table~\ref{tab:paper-structure}. Additionally, we include an online
repository containing all code used to generate the results in this
paper; see \texttt{\url{https://github.com/connor-duffin/sswe}}. \\

\begin{tablehere}
  \centering
  \begin{tabularx}{\linewidth}{rX}
    \hline
    Section & Contents \\
    \hline
    \ref{sec:modelling} & Physical model, \gls*{GP} priors, discretisation, algorithms. \\
    \ref{sec:experimental-setup} & Data generation, noise level, hyperparameters, prior distribution. \\
    \ref{sec:intro-examples} & Posterior distribution: introductory examples RMSE. \\
    \ref{sec:obs-frequency} & Posterior distribution: analysis of spatiotemporal observation frequency. \\
    \ref{sec:parameter-misspecification} & Posterior distribution: bathymetry and viscosity misspecification. \\
    \ref{sec:linearisation} & Posterior distribution: linear model results with bathymetry and viscosity misspecification. \\
    \ref{sec:conclusion} & Discussion and conclusion. \\
    \hline
    Code & \texttt{\url{https://github.com/connor-duffin/sswe}} \\
    \hline
  \end{tabularx}
  \caption{Quick-reference paper structure.}
  \label{tab:paper-structure}
\end{tablehere}

\section{Physical-statistical model}
\label{sec:modelling}

\begin{figure}[H]
  \centering
  \includegraphics[width=0.45\textwidth]{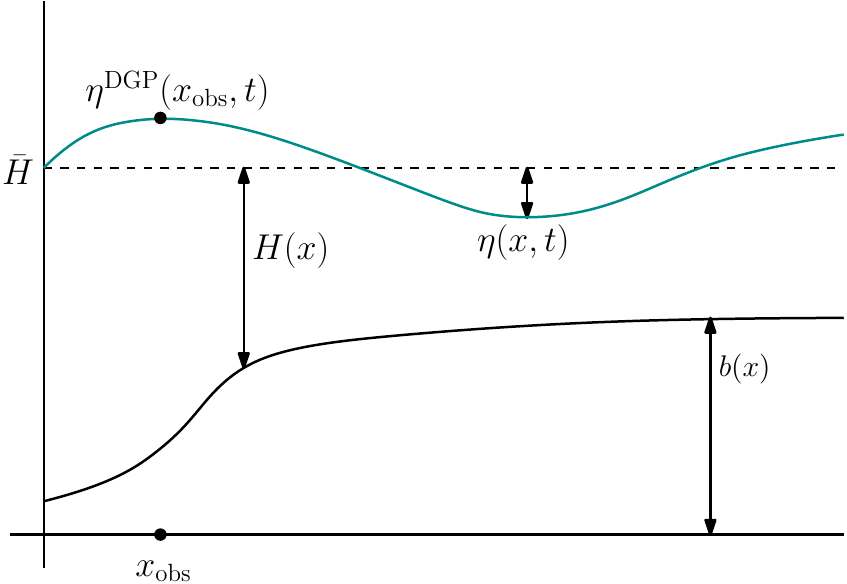}
  \caption{Illustration of the bathymetry $b(x)$, mean fluid depth $H(x)$,
    surface height perturbation $\eta(x, t)$, and the observation system.}
  \label{fig:observation-system}
\end{figure}

For our example system we use the one-dimensional \gls*{SWE}. The
\gls*{SWE} are derived from the two-dimensional incompressible
Navier-Stokes equations through integrating over the vertical
direction~\citep{cushman-roisin2011introduction}. In this work we also assume that
the single-layer flow is irrotational.
What results is a coupled \gls*{PDE} system consisting of state variables $(u, \eta) \in
\bR^2$, with $u := u(x, t)$, the velocity field, and
$\eta := \eta(x, t)$, the surface height, for spatial variable $x$ and time variable
$t$. Our model is that of an idealised, tidally forced flow into an
inlet with a spatial domain of length $10$ \si{\kilo\metre}. The model
employed is thus:
\begin{equation}
  \label{eq:swe-deterministic}
  \begin{cases}
    u_t + u u_x - \nu u_{xx} + g \eta_x = 0, & x \in [0, 10000], \\
    \eta_t + \left((H + \eta) u \right)_{x} = 0, & x \in [0, 10000], \\
    u(10000, t) = 0, \; \eta(0, t) = \tau(t).
  \end{cases}
\end{equation}
The tidal forcing is
\begin{equation}
  \label{eq:tidal-bc}
  \tau(t) := 2 \left(1 + \cos \left( \frac{4 \pi t}{86400} \right) \right).
\end{equation}
The mean surface height, $H(x)$ implies the topography $b(x)$ of the
solution domain. In our setting therefore we set $H(x) = \bar{H} - b(x)$,
with $\bar{H} = 30$. The topography $b(x)$ is a gradual sloping shore
with a horizontal displacement parameter $s$:
\begin{equation}
  \label{eq:topography}
  b(x) = 5 \left(1 + \tanh \left( \frac{x - s}{2000} \right) \right).
\end{equation}
The fluid starts at rest, $u(x, 0) \equiv 0$, $\eta(x, 0) = 0$, and
the model is run up to time $t = 12$ \si{\hour}. An illustration of
these functions is shown in Figure~\ref{fig:observation-system}.

We also consider the linearised version of
\eqref{eq:swe-deterministic}, which ignores second-order terms and
assumes that $\eta \ll H$. This gives
\begin{equation}
  \label{eq:swe-linear-deterministic}
  \begin{cases}
    u_t - \nu u_{xx} + g \eta_x= 0, & x \in [0, 10000], \\
    \eta_t + \left(H u \right)_{x} = 0, & x \in [0, 10000], \\
    u(10000, t) = 0, \; \eta(0, t) = \tau(t).
  \end{cases}
\end{equation}
Initial conditions, bathymetry, and tidal forcing are the same as
those for the fully nonlinear system.

To reconcile the model with observed data we begin by introducing
uncertainty into the governing equations (i.e.,
\eqref{eq:swe-deterministic} or \eqref{eq:swe-linear-deterministic})
through additive \gls*{GP} forcing, following \gls*{statFEM}. This
derives a prior distribution which forms the reference measure for
posterior inference. For the nonlinear case this is
\begin{equation}
  \label{eq:swe-stochastic}
  \begin{cases}
    u_t + u u_x - \nu u_{xx} + g \eta_x = \xi_u, & x \in [0, 10000], \\
    \eta_t + \left((H + \eta) u \right)_{x} = \xi_\eta, & x \in [0, 10000], \\
    u(10000, t) = 0, \; \eta(0, t) = \tau(t).
  \end{cases}
\end{equation}
The linear case follows similarly and is detailed in Appendix~\ref{sec:linear-model-derivation}.
The \textit{a priori} uncorrelated \gls*{GP} forcing terms $\xi_u$ and
$\xi_\eta$ are given by
\[
  \begin{pmatrix}
    \xi_u \\ \xi_\eta
  \end{pmatrix} \sim
  \GP\left(
  \begin{pmatrix}
    0 \\ 0
  \end{pmatrix},
  \delta(t - t')
  \begin{bmatrix}
    k_u(\cdot, \cdot) & 0 \\
    0 & k_\eta(\cdot, \cdot)
  \end{bmatrix}
  \right).
\]
The kernels $k_u(\cdot, \cdot)$ and $k_\eta(\cdot, \cdot)$ have
hyperparameters $\bm{\Theta}$ which in this work are fixed and
known. Estimation methods are
available~\citep[see, e.g.,][]{williams2006gaussian}, but we choose to
fix parameters for consistency across comparisons, as, in this work, we
are interested only in the posterior \gls*{statFEM} filtering
inference --- not in joint filtering and hyperparameter inference. We
use the squared-exponential kernel, given by
$k(\vx, \vx') = \rho^2 \exp(-\lVert \vx - \vx' \rVert^2 / (2 \ell^2))$,
which we notationally subscript to represent the individual component
kernels $k_u(\cdot, \cdot)$ and $k_\eta(\cdot, \cdot)$, with hyperparameters
$\bm{\Theta} = \{\bm{\Theta}_u, \bm{\Theta}_\eta\}
= \{\rho_u, \ell_u, \rho_\eta, \ell_\eta \}$.

Discretisation of this system now proceeds via the \gls*{FEM} to give
a finite-dimensional approximation to the prior. To do so we use the
discretisation of~\citet{jacobs2015firedrakefluids}. We use a
uniform mesh $\cD_h \subseteq \cD$ with vertices
$\{x_j\}_{j = 1}^{n_v}$; the subinterval length is $h$. We use the
$P2$-$P1$ element pair to discretise the state, giving the basis
function expansions of
$u(x, t) \approx u_h(x, t) = \sum_{i = 1}^{n_u} u_{i}(t) \phi_i(x)$,
$\eta(x, t) \approx \eta_h(x, t) = \sum_{i = 1}^{n_\eta} \eta_{i}(t) \psi_i(x)$.
The span of the basis functions $\{\phi_i\}_{i = 1}^{n_u}$ and $\{\psi_i\}_{i = 1}^{n_\eta}$
defines the \gls*{FEM} trial and test spaces for the velocity and
surface height perturbations, respectively. The weak form of
\eqref{eq:swe-stochastic}~is given by multiplying by testing functions
$(v_u, v_\eta)$ and integrating over the spatial domain $\cD$
\begin{gather*}
  \langle u_t, v_u \rangle
  + \langle u u_x \rangle
  + \nu \langle u_{x}, v_{u, x}\rangle
  + g \langle \eta_x, v_u \rangle
  = \langle \xi_u, v_u \rangle,  \\
  \langle \eta_t, v_\eta \rangle
  + \langle \left((H + \eta) u \right)_{x}, v_\eta \rangle
  = \langle \xi_\eta, v_\eta \rangle, 
\end{gather*}
where $\langle f, g \rangle = \int_\cD f g \, \dee x$. Substituting
the finite-dimensional \gls*{FEM} approximations to both the trial and
test functions gives differential equations over the \gls*{FEM} coefficients
$\vu = (u_1, \ldots, u_{n_u})$, $\bm{\eta} = (\eta_1, \ldots, \eta_{n_\eta})$:
\begin{gather*}
  \mM_u \frac{\partial \vu}{\partial t}
  + \cF_u(\vu)
  + \nu \mA \vu 
  + g \mB \bm{\eta}
  = \bm{\xi}_u, \\
  \mM_\eta \frac{\partial \bm{\eta}}{\partial t}
  + \cF_\eta(\vu, \bm{\eta})
  = \bm{\xi}_\eta,
\end{gather*}
where $\mM_{u, ji} = \langle \phi_i, \phi_j \rangle$,
$\mA_{ji} = \langle \phi_{i, x}, \phi_{j, x} \rangle$,
$\mB_{ji} = \langle \psi_{i, x}, \phi_{j} \rangle$, and
$\cF_u(\cdot)$, $\cF_\eta(\cdot, \cdot)$ are functions which result
from discretising the nonlinear operators.

The \gls*{FEM} discretised \gls*{GP} forcing terms,
$\bm{\xi}_u$ and $\bm{\xi}_\eta$, are given by the approximation
$\bm{\xi} \sim \NPDF(\vzero, \mM \mK
\mM^\top)$, where $\mK_{ij} = k(x_i, x_j)$, for nodal $x_i$,
$x_j$~\citep{duffin2021statistical} (note omitted subscripts are
for readability). This approximation uses
different mass matrices $\mM$ across components due to the components
$u$ and $\eta$ using different basis functions for the \gls*{FEM}
approximation. Thus we have, jointly,
$(\bm{\xi}_u, \bm{\xi}_\eta) \sim \GP(\vzero, \delta(t - t') \mG)$,
where $\mG$ has the block-diagonal structure:
\[
  \mG =
  \begin{bmatrix}
    \mM_u \mK_u \mM_u^\top & \vzero \\
    \vzero & \mM_\eta \mK_\eta \mM_\eta^\top
  \end{bmatrix}.
\]
A low-rank approximation is required, in order to run our filtering
methodology~\citep{duffin2022lowrank}. To get a low-rank approximation
of this covariance matrix we make use of the block structure. Using
the factorisation $\mG = \mG^{1/2} \mG^{\top/2}$, we approximate
\[
  \mG^{1/2}
  \approx
  \begin{bmatrix}
    \mM_u \mK_u^{1/2} & \vzero \\
    \vzero & \mM_\eta \mK_\eta^{1/2}
  \end{bmatrix},
\]
where $\mK_u^{1/2} \in \bR^{n_u \times q_u}$ and
$\mK_\eta^{1/2} \in \bR^{n_\eta \times q_\eta}$, for $q_u \ll n_u$, $q_\eta \ll n_\eta$.
The block-structured low-rank approximation gives
$(\bm{\xi}_u, \bm{\xi}_\eta)^\top \sim \GP(\vzero, \delta(t - t') \mG^{1/2} \mG^{\top/2})$,
where $\mG^{1/2} \in \bR^{(n_u + n_\eta) \times (q_u + q_\eta)}$.
Approximations can be computed through e.g., GPU computing~\citep{charlier2021kernel}
or Nystr\"om approximation~\citep{williams2001using}, but in
this work we use the Hilbert-\gls*{GP} approach
of~\citet{solin2020hilbert}. This enforces that the additive \glspl*{GP} should be
zero on the boundaries. It was found empirically that our \gls*{GP}
approximations needed to respect zero boundary conditions or otherwise
the posterior covariance would be overly uncertain on the edges of the
domain, leading to poor numerical approximation of the covariance.

To discretise the dynamics in time we use the
$\theta$-method~\citep{hairer1993solving} for stability. Letting
$\vu^n := \vu(n \Dt)$ and
$\bm{\eta}^n := \bm{\eta}(n \Dt)$, the time-discretised
stochastic dynamics are
\begin{multline*}
  \mM_u \frac{\vu^{n} - \vu^{n - 1}}{\Dt} + \cF_u(\vu^{n - \theta}) \\
  + \nu \mA \vu^{n - \theta} + g \mB \bm{\eta}^{n - \theta} = \frac{1}{\sqrt{\Dt}} \bm{\xi}_u^{n - 1}, \\
  \mM_\eta \frac{\bm{\eta}^n - \bm{\eta}^{n - 1}}{\Dt} + \cF_\eta(\vu, \bm{\eta}^{n - \theta}) = \frac{1}{\sqrt{\Dt}} \bm{\xi}_\eta^{n - 1},
\end{multline*}
where $\vu^{n - \theta} := \theta \vu^{n} + (1 - \theta) \vu^{n - 1}$
(similarly for $\bm{\eta}$), for $\theta \in [0, 1]$. The initial conditions $(\vu^0,
\bm{\eta}^0)$ are known so we begin by solving for $(\vu^1, \bm{\eta}^1)$. Running the scheme
gives the entire set of states $\{(\vu^n, \bm{\eta}^n)\}_{n = 0}^N$ so that $N\Dt = T$.

Observations may also be arriving at particular timepoints, and we
want to condition on these observations to get an \apost{} estimate of
the state. We assume that the time between observations is
$k \Dt$, for some integer $k \geq 1$, giving the observations
as $\vy_m := \vy(m k \Dt)$, and the total set of observations
$\{\vy_m\}_{m = 1}^M$ --- the initial state $(\vu^n, \bm{\eta}^n)$ is
always assumed known. This ensures
$\{m k \Dt\}_{m = 1}^M \subset \{n \Dt \}_{n = 0}^N$ and thus $M \leq N$.

The joint dynamics and observation model is to take $k$ model steps, thus for $n = (m - 1) k, \ldots, mk$
we predict using the model:
\begin{equation}
  \begin{gathered}
    \label{eq:full-swe-discretised}
    \mM_u \frac{\vu^{n} - \vu^{n - 1}}{\Dt} + \cF_u(\vu^{n - \theta}) \\
    \qquad \qquad \qquad \quad + \nu \mA \vu^{n - \theta} + g \mB \bm{\eta}^{n - \theta} = \frac{1}{\sqrt{\Dt}}\bm{\xi}_u^{n - 1}, \\
    \mM_\eta \frac{\bm{\eta}^n - \bm{\eta}^{n - 1}}{\Dt} + \cF_\eta(\vu, \bm{\eta}^{n - \theta}) = \frac{1}{\sqrt{\Dt}} \bm{\xi}_\eta^{n - 1}.
  \end{gathered}
\end{equation}
We abbreviate this by writing the l.h.s. of \eqref{eq:full-swe-discretised}
as $\cM : \bR^{(n_u + n_\eta) \times 2} \to \bR^{n_u + n_\eta}$, with Jacobian matrix
$\mJ_n$.

At the observation timepoint $mk \Dt$ we condition on the data
\begin{equation*}
  \vy_m = \mH (\vu_m, \bm{\eta}_m) + \vr_m, 
\end{equation*}
where $(\vu_m, \bm{\eta}_m) = (\vu^{mk}, \bm{\eta}^{mk})$
and $\vr_m \sim \NPDF(\vzero, \mR)$. The linear observation
operator $\mH: \bR^{n_u + n_\eta} \to \bR^{n_y}$ is known \aprio{}.
In this work, it is given by the \gls*{FEM} polynomial interpolants. To condition on
the observations we use the \gls*{LR-ExKF} --- a
recursive two-step scheme consisting of prediction and update
steps. At timesteps which are not observed only the model prediction
steps are completed. This computes the approximation
$p(\vu_m, \bm{\eta}_m \given \vy_{1:m}, \bm{\Theta}, \nu, c) \sim \NPDF(\bm{\mu}_m, \mL_m \mL_m^\top)$,
a multivariate Gaussian over the concatenation of $(\vu_m, \bm{\eta}_m)$, an $n_u + n_\eta$
dimensional object. For a rank-$q$ approximation to the covariance matrix we thus have
$\mL_m \in \bR^{(n_u + n_\eta) \times q}$. For a single
prediction-update cycle, the algorithm is shown in Algorithm~\ref{alg:lr-exkf}.

\begin{algorithm*}[t]
  \begin{algorithmic}
    \Require $\bm{\mu}_{m - 1}$,  $\mL_{m - 1}$, such that $(\vu_{m - 1}, \bm{\eta}_{m - 1} \given \bm{\Theta}, \nu, c) \sim \NPDF(\bm{\mu}_{m - 1}, \mL_{m - 1} \mL_{m - 1}^\top)$.
    \State Let $\bm{\mu}^n \equiv \bm{\mu}_{m - 1}$, $\mL^n \equiv \mL_{m - 1}$.
    \For {$(m - 1) k < n \leq mk$}
    \State Solve $\cM(\bm{\mu}^{n}, \bm{\mu}^{n - 1}) = \vzero$, for $\bm{\mu}^n$.
    \State $\tilde\mL^{n} = \left[\mJ_{n}^{-1} \mJ_{n - 1} \mL_{n - 1}, \; \mJ_{n}^{-1} \mG^{1/2} \right]$.
    \State Eigendecomposition: $\mV_n \bm{\Sigma}_n \mV_n^\top = ({\tilde \mL}^{n})^\top \tilde\mL^{n}$.
    \State $\mL^n = \tilde\mL^{n} [\mV]_{:, 1:q}$.
    \EndFor
    \State $\bm{\mu}^{n} \gets \bm{\mu}^{n}
        + \mL^n (\mH \mL^n)^\top (\mH \mL^n (\mH \mL^n)^\top + \sigma^2 \mI)^{-1} (\vy_m - \mH \bm{\mu}^{n})$.
    \State Cholesky decomposition: $\mR_n \mR_n^\top
        = \mI - (\mH \mL^n)^\top (\mH \mL^n (\mH \mL^n)^\top + \sigma^2 \mI)^{-1} \mH \mL^n$.
    \State $\mL^{n} \gets \mL^{n} \mR_{n}$.
    \State \Return $\bm{\mu}_{m} \equiv \bm{\mu}^n$, $\mL_m \equiv \mL^{n}$.
  \end{algorithmic}
  \caption{Prediction-update cycle of the \acrshort*{LR-ExKF} algorithm (rank $q$).}
  \label{alg:lr-exkf}
\end{algorithm*}

\begin{figure*}[t]
  \centering
  \includegraphics[width=\textwidth]{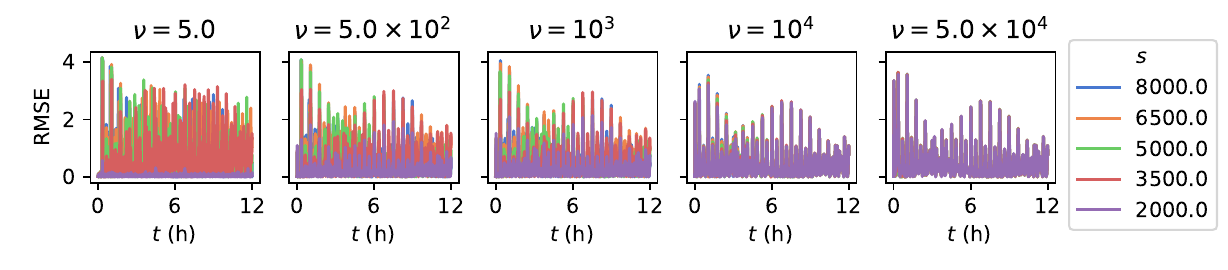}
  \caption{Prior \gls*{statFEM} RMSE: computed for models with
    $s \in \{2000, 3500, 5000, 6500, 8000\}$ and
    $\nu \in \{5, 500, 10^3, 10^4, 5 \times 10^4\}$, every $30$ s.}
  \label{fig:prior-nu-s-rmse}
\end{figure*}

\section{Experimental setup}
\label{sec:experimental-setup}

To generate the synthetic dataset, $\{\vy_m \}_{m = 1}^M$, we use the
\gls*{FEM} discretisation of the fully nonlinear \gls*{SWE} (of
Equation~\eqref{eq:swe-deterministic}) as detailed above for the
\gls*{statFEM} model. That is, we use the same $P2$-$P1$ basis
function pairs to give the \gls*{FEM} discretised approximations
$(u_h^{\mathrm{DGP}}(x, t), \eta_h^{\mathrm{DGP}}(x, t))$. These are computed using
a uniform mesh with $n_v = 500$ elements ($h = 20$ \si{\metre}), and
timesteps of size $\Dt = 1$ \si{\second}. We set $\theta =
0.6$. Observations are given by
\begin{multline*}
  \vy_m = \bm{\eta}_m^{\mathrm{obs}} + \vr_m, \\
  \bm{\eta}_m^{\mathrm{obs}} := \left(
    \eta_h^{\mathrm{DGP}}(x_1^{\mathrm{obs}}, mk \Dt), \ldots,
    \eta_h^{\mathrm{DGP}}(x_{n_y}^{\mathrm{obs}}, mk \Dt)
  \right)^\top,
\end{multline*}
where the i.i.d. noise is $\vr_m \sim \NPDF(\vzero, \sigma^2 \mI)$,
with $\sigma = 5 \times 10^{-2}$. This data is generated with $\nu =
1$, and the shore position is $s = 2000$. We take $n_y$ observations per observed
time point, the locations of which are uniformly spaced between in the
interval $[1000, 2000]$ \si{\metre}. Note that when $n_y = 1$ this corresponds
to observing at $x_{\mathrm{obs}} = 1000$ \si{\metre}.

Our experiments compare results \aprio{} with results \apost{} under different model
configurations to those used to generate the data. The posterior distribution
is given by $p(\vu_m, \bm{\eta}_m \given \vy_{1:m}, \bm{\Theta}, \sigma, \nu, c)$,
which we compute an approximation to using the \gls*{LR-ExKF}. The posterior $(u_h, \eta_h)$
is computed using the same numerical settings as for the data. The observation
operator $\mH$ is defined via
  \begin{multline*}
    \mH(\vu_m, \bm{\eta}_m) := \\
    \left(\eta_h(x_1^{\mathrm{obs}}, mk \Dt), \ldots,
    \eta_h(x_{n_y}^{\mathrm{obs}}, mk \Dt) \right)^\top,
\end{multline*}
and we assume that the noise level $\sigma$ is known, simulating the
scenario of known measurement device error. We compute the posterior
distribution across a
Cartesian product of the different input parameters, with $k \in \{1,
30, 60, 120, 180\}$, $n_y \in \{1, 2, 5\}$, $s \in \{2000, 3500, 5000,
6500, 8000\}$, and $\nu \in \{5, 500, 1000, 10000, 50000\}$. Across the nonlinear
and linear models this gives $750$ different configurations. Note that we do not
estimate the \gls*{statFEM} posterior for $\nu = 1$ due to
numerical instabilities when computing the posterior covariance,
however results were similar to that with $\nu = 5$, which is reported here.
 For the stochastic \gls*{GP} forcing, we set
$\ell_u = \ell_\eta = 1000$, $\rho_u = 0$, and
$\rho_\eta = 2 \times 10^{-3}$. The magnitudes of $\rho_u$ and $\rho_\eta$ are
chosen to balance between accurate UQ when estimating a well-specified model, and
adequate uncertainty when estimating a poorly specified model.

To get a feel for model performance \aprio{} --- and hence the severity of model
misspecification --- we estimate the prior distribution for the nonlinear model,
$p(\vu_m, \bm{\eta}_m \given \nu, s, \bm{\Theta}) \sim \NPDF(\bm{\mu}_m, \mL_m \mL_m^\top)$,
across the grid of $s$ and $\nu$ values. This is done through running
the filter with the prediction steps only, for all timesteps. To
compare with the data we compute the \gls*{RMSE}. The \gls*{RMSE} is
\begin{equation}
  \label{eq:rmse-def}
  \RMSE_m = \frac{\lVert \vy_m - \mH \bm{\mu}_m \rVert_2}{\sqrt{n_y}},
\end{equation}
where $\lVert \cdot \rVert_2$ is the Euclidean $l^2$ norm.  Results
are shown in Figure~\ref{fig:prior-nu-s-rmse}. For $\nu = 5$ there is a clear
stratification between the well-specified $s = 2000$ model and the others
which are misspecified. The errors in these models appear to lack the consistent
periodicity that models with larger $\nu$ see. In these cases we see that there is a
consistently large error across each model with no synchronicity across the systems.
The stratification between these models becomes less apparent as $\nu$
increases up to $5\times 10^4$, a result of the dissipative effects
dominating the dynamics. This leads to models with different
$s$ performing similarly as the wave profiles dissipate the energy input
from the tidal forcing.

There emerges a periodicity across the solutions as $\nu$ increases,
thought to be due to the tidal forcing. We see that there is a sharp
increase in early times, then a similar increase approximately in the
middle of the time domain. This increase is thought to be due to the
cycle of the forcing starting to ``swing down'' into the lower cycle
of the tidal forcing. We note also that there are similar timescales
in the error dynamics and no models appear to dissipate to equilibrium ---
again due to the oscillatory forcing.

% For large $\nu$ models, with $\nu \geq 10^4$, we see that there is
% less stratification and the model is dominated by the viscous
% effects. In each case the models rapidly reach $\RRMSE \approx 1$,
% which coincides with the dissipation of the surface heights to $\eta_h \approx 0$.
% Mild stratification is present, across the $c$ values, for
% $\nu = 0.1$, which is otherwise not seen when we increase $\nu =
% 1$. In this case we see that the dissipation to steady-state happens
% near-instantly after the model is initialised. In general these plots
% demonstrate that there are different timescales that result in
% mismatch with the dynamical system. When $c$ is misspecified
% discrepancies arise after $O(100)$ s.
% For $\nu$ the misspecification
% results, in the most severe case, in near instant dissipation and is
% thus much more rapid.

% We see the interesting
% phenomena that if the bump is misspecified $5$ m to the left ($c =
% 5$), errors are much larger than $5$ m to the right ($c = 15$). This
% is thought to be due to the proximity of boundary interactions that
% result when $c = 5$. Note also the symmetry that arises with the
% errors being similar for $c = 5$ and $c = 20$, which both have a
% minimum absolute boundary distance of $5$ m.

\section{Results}
\label{sec:results}

In this section we analyse the posterior results. First, we conduct a
preliminary analysis of the model posteriors, to give intuition
on how our chosen metrics relate to the posterior distribution. Next,
we analyse how the observation frequencies $k$ and $n_y$ effect the posterior
distribution in the face of misspecification. We then study the case of
joint viscosity-bathymetry misspecification, and then conclude with the
analysis of the linearised model, also under joint
viscosity-bathymetry misspecification.

\begin{figure*}[t]
  \centering
  \begin{subfigure}[t]{0.7\textwidth}
    \centering
    \includegraphics[width=\textwidth]{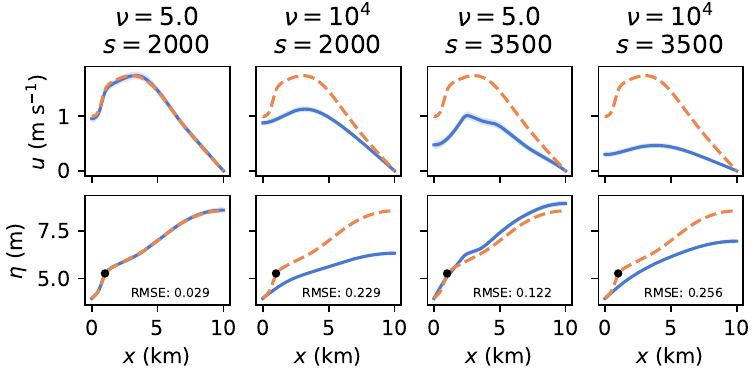}
    \caption{Observations, posterior means and $0.95$ probability
      intervals, at time $t = 11.67$ \si{\hour}. Average RMSE is shown
      within each plot, and the truth is shown in
      \textcolor{tab_orange}{orange}~(\textcolor{tab_orange}{\rule[1.5pt]{0.15cm}{1pt} \rule[1.5pt]{0.15cm}{1pt}}).}
    \label{fig:intro-example-means-vars}
  \end{subfigure}~
  \begin{subfigure}[t]{0.25\textwidth}
    \centering
    \includegraphics[width=\textwidth]{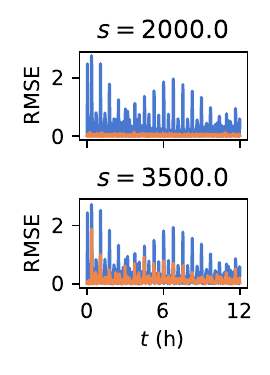}
    \caption{RMSE: $\nu = 5$ shown in
      \textcolor{tab_orange}{orange}~(\textcolor{tab_orange}{\rule[1.5pt]{0.4cm}{1pt}})
      and $\nu = 10^4$ shown in \textcolor{tab_blue}{blue}~(\textcolor{tab_blue}{\rule[1.5pt]{0.4cm}{1pt}}).}
    \label{fig:intro-example-rmse}
  \end{subfigure}

  \caption{Posterior means and variances
    (\subref{fig:intro-example-means-vars}), and posterior
    \acrshort*{RMSE} values (\subref{fig:intro-example-rmse}) for the
    nonlinear models with $s \in \{2000, 3500\}$
    \si{\metre}, $\nu \in \{5, 10^4\}$ \si{\metre^2\per\second}.}
    \label{fig:intro-example}
\end{figure*}

\begin{figure*}[htbp]
  \centering
  \begin{subfigure}[t]{\textwidth}
    \centering
    \includegraphics[width=\textwidth]{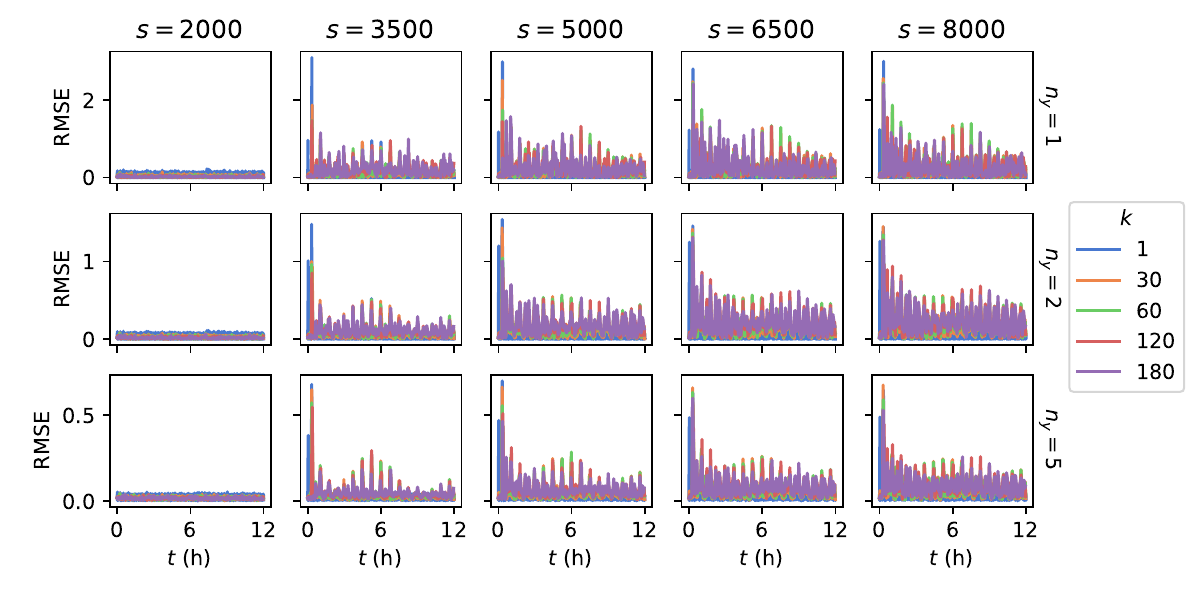}
    \caption{\gls*{RMSE} values across time, across different $s$, $n_y$, and $k$ ($\nu = 5$).}
    \label{fig:rmse-post-s-nt-skip-nx-obs}
  \end{subfigure}
  \begin{subfigure}[t]{\textwidth}
    \centering
    \includegraphics[width=\textwidth]{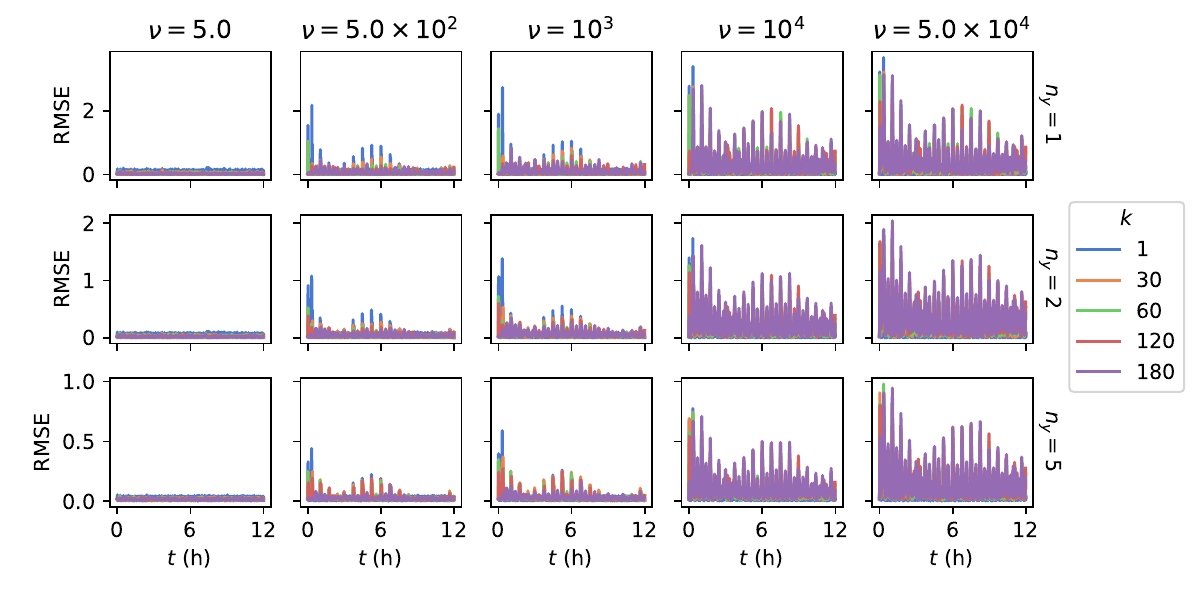}
  \caption{\gls*{RMSE} values across time as $\nu$, $n_y$, and $k$ are varied ($s = 2000$).}
  \label{fig:rmse-post-nu-nt-skip-nx-obs}
  \end{subfigure}
  \caption{Observation frequency: posterior \gls*{RMSE} values, across time, for the models as
    observation frequency is varied.}
  \label{fig:nt-skip-nx-obs-rmse}
\end{figure*}

\subsection{Preliminary analysis of posterior distributions}
\label{sec:intro-examples}

To get a feel for the posterior results we now describe the results
for four models. Each have observations arriving every $k = 30$
timesteps (every $30$ \si{\second}). We run the nonlinear model
with $s \in \{2000, 3500\}$ and $\nu \in \{5, 10^4\}$.

At time $t = 11.67$ \si{\hour} we have plotted
the posterior means and variances in
Figure~\ref{fig:intro-example-means-vars}. The well specified model
captures the more complex dynamical behaviour well with a notable improvement in
the estimation of the velocity fields, in comparison to the other
models. The more damped models, with $\nu = 10^4$, appear unsurprisingly
to underestimate the data at this observation point.
Due to the right-shifted bathymetry an increase in velocity is seen to
the right of the observation location when $s = 3500$, $\nu = 5$.
The velocity fields are all underestimated, with a notably
poor-performing case with $s = 3500$ \si{\metre} and $\nu = 10^4$.
In this case the data (observed only on the surface height
perturbation) can only correct for so much, and the dynamics must also
be appropriately specified in order for the model to be accurate. We
also see that the uncertainty on $\eta$ has given rise to uncertainty
in $u$ following intuition; \apost{} it is seen that the unobserved
velocity components have increased uncertainty.

As introduced above, to quantitatively compare performance we use the
\gls*{RMSE}. For the models introduced above the average values of
these, across all time, are shown within the second row in
Figure~\ref{fig:intro-example-means-vars}. Across the variations in
\gls*{RMSE} there is a qualitative stratification which is especially
apparent on the unobserved velocity components. The \glspl*{RMSE} are
plotted across time in Figure~\ref{fig:intro-example-rmse}; similar
stratification is seen to that in
Figure~\ref{fig:intro-example-means-vars}. Variation is seen across
the models as data is conditioned on; this is most clearly observed
with the poorly performing high-viscosity models. The low-viscosity
model with $\nu = 5$ performs well. The mildly-misspecified $\{\nu =
5, s = 3500\}$ performs moderately well and improves
upon the prior (see Figure~\ref{fig:prior-nu-s-rmse}).

\subsection{Investigating observation frequency}
\label{sec:obs-frequency}

In the second simulation study we study the model performance as we
vary the observation frequency in space and time, taking $n_y \in \{1, 2, 5\}$ and
$k \in \{1, 30, 60, 120, 180\}$, whilst also
varying the topography and viscosity.

First, we look at the case of a well-specified viscosity ($\nu = 5$), 
with misspecified bathymetry, with $s \in \{2000, 3500, 5000, 6500, 8000\}$.
In Figure~\ref{fig:rmse-post-s-nt-skip-nx-obs} we plot the \gls*{RMSE}
values over all the observed timepoints, for each model. Increasing
$n_y$ decreases the model error across all models. Similar reductions in
the \gls*{RMSE} are not seen with the increase of $k$. Whilst there are
improvements, especially for all $k = 1$, and $n_y = 1$, it is seen
otherwise that the observation frequency $k$ does not have the same
drastic effect.

\begin{figure}[H]
  \centering
  \includegraphics[width=0.4\textwidth]{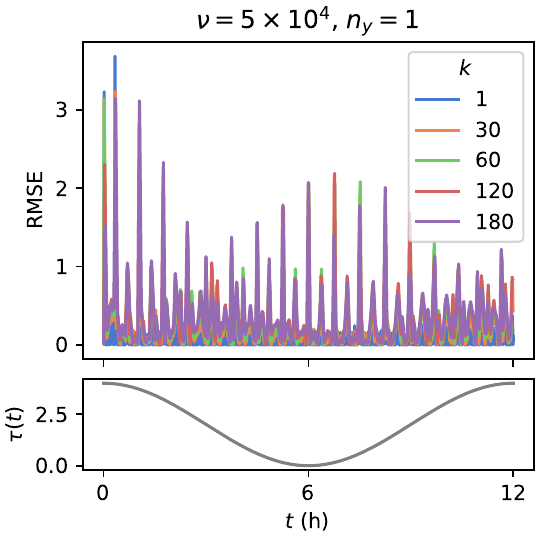}
  \caption{RMSE and tidal forcing $\tau(t)$ with $\nu = 5 \times 10^4$
    and $n_y = 1$. Errors increase when the forcing is at its minimum.}
  \label{fig:rmse-post-nu-single-tidal}
\end{figure}

We next look at the case of a well-specified $s = 2000$ and a variable
viscosity $\nu \in \{5, 500, 10^3, 10^4, 5 \times 10^4 \}$. Results
when varying $\nu$ are shown in
Figure~\ref{fig:rmse-post-nu-nt-skip-nx-obs}. For $\nu \leq 10^3$ we
see that the models perform relatively well; conditioning on data
ensures that the misspecification induced through the viscosity is
corrected for. As $\nu$ increases we see the regular increase in
error, through the middle of simulation time (at $t \approx
6$~\si{\hour}). Whilst the \gls*{RMSE} values vary magnitude-wise,
this regular quasi-periodic structure emerges across each of the
models. This approximately corresponds to the tidal forcing $\tau(t)$
hitting its minimum through the simulation
(Figure~\ref{fig:rmse-post-nu-single-tidal}). Errors decrease as this
forcing begins to increase once again.

Increasing the observation density in space again results in a marked
improvement in model discrepancy. As previous with $k = 1$ this
results in the most notable improvement in the \gls*{RMSE}, with
mild improvements for $k \geq 30$. As $\nu \geq 10^4$, there is no
visual distinction between the models with such high
viscosities. Error due to the topography appears to result in a
greater degree of stratification between each of the models. This is
unsurprising as whilst misspecifying the viscosity leads to mismatch,
beyond $\nu = 10^4$ the viscous effects dominate the flow, resulting
in similar behaviour.

% table
\begin{table*}[t]
  \centering
  \begin{tabular}{rlllll}
\hline
    & 1              & 30             & 60             & 120            & 180            \\
\hline
  1 & 0.0543 (0.080) & 0.1222 (0.123) & 0.1447 (0.128) & 0.1792 (0.153) & 0.1944 (0.177) \\
  2 & 0.0417 (0.048) & 0.0720 (0.073) & 0.0857 (0.072) & 0.1019 (0.080) & 0.1127 (0.079) \\
  5 & 0.0266 (0.026) & 0.0356 (0.040) & 0.0389 (0.042) & 0.0438 (0.045) & 0.0452 (0.027) \\
\hline
\end{tabular}
  \caption{Mean RMSE values and standard deviations (in parentheses)
    for the nonlinear statFEM SWE model, as the observation frequency is
    changed, with mild misspecification: $s = 3500$ and $\nu = 5$.}
  \label{tab:s-misspec-rmse-nt-skip-nx-obs}
\end{table*}

% \begin{table*}[t]
%   \centering
%   \input{figures/nt-skip-nu-table.tex}
%   \caption{$\overline{\RRMSE}$ for the nonlinear statFEM SWE model, with
%     $c = 10$, as $\nu$ and $k$ are varied. }
%   \label{tab:nt-skip-nu-table}
% \end{table*}

For a single instance of ``mild misspecification'' with $s = 3500$ and
$\nu = 5$, the empirical means and standard deviations (computed
across time) of their \glspl*{RMSE} are shown in
Table~\ref{tab:s-misspec-rmse-nt-skip-nx-obs}. As more spatial
locations are observed, the frequency of observations in time has less
of an effect on the accuracy of the model. When observing $n_y = 5$
locations, small increases in the \gls*{RMSE} are seen with less
frequent observations in time. These increases are notably larger
when taking $n_y = 1$. An interesting result is that increasing
$n_y$ from $1$ to $5$ results in improved performance, even when
observing every $180$ \si{\second}. The inclusion of additional
spatial measurement locations results in a dramatic improvement in the
performance of the model. This is thought to be due to the fact that
the flow in this case has a long wavelength --- incorporating data
over a larger spatial domain therefore has a more corrective effect on
the model as it is now observed over a set of spatiotemporal
locations.

\begin{figure*}[t]
\centering
\includegraphics[width=\textwidth]{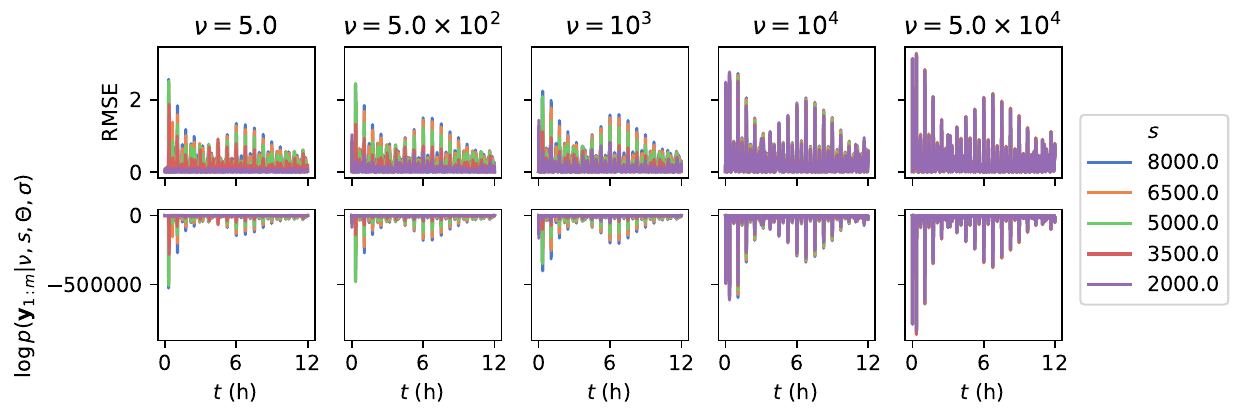}
\caption{Posterior bathymetry-viscosity
    misspecification. \gls*{RMSE} values as $\nu$ and $c$ are varied
    (top), and negative log-likelihood values
    (bottom), for fixed observation interval $k = 100$.}
\label{fig:nu-s-rmse-nlml-layout}
\end{figure*}

\begin{figure*}[t]
  \centering
  \includegraphics[width=0.8\textwidth]{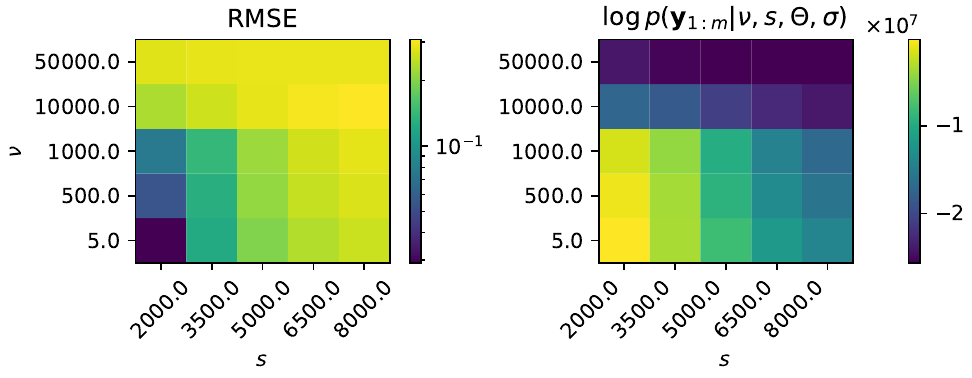}
  \caption{Color table of \gls*{RMSE} and negative log-likelihood results as
    $s$ and $\nu$ are varied. For numeric comparisons with the above
    figures, the \gls*{RMSE} is averaged across all observation times.}
  \label{fig:nu-s-rmse-lml-table-color}
\end{figure*}

\begin{table*}[t]
  \centering
  \begin{tabular}{rlllll}
\hline
       & 2000.0                          & 3500.0                          & 5000.0                          & 6500.0                          & 8000.0                          \\
\hline
     5 & $\mathbf{-1.541 \times 10^{5}}$ & $\mathbf{-3.453 \times 10^{6}}$ & $\mathbf{-8.174 \times 10^{6}}$ & $\mathbf{-1.180 \times 10^{7}}$ & $\mathbf{-1.406 \times 10^{7}}$ \\
   500 & $-8.184 \times 10^{5}$          & $-3.599 \times 10^{6}$          & $-9.216 \times 10^{6}$          & $-1.346 \times 10^{7}$          & $-1.582 \times 10^{7}$          \\
  1000 & $-1.795 \times 10^{6}$          & $-4.249 \times 10^{6}$          & $-9.866 \times 10^{6}$          & $-1.441 \times 10^{7}$          & $-1.701 \times 10^{7}$          \\
 10000 & $-1.736 \times 10^{7}$          & $-1.855 \times 10^{7}$          & $-2.091 \times 10^{7}$          & $-2.284 \times 10^{7}$          & $-2.403 \times 10^{7}$          \\
 50000 & $-2.409 \times 10^{7}$          & $-2.541 \times 10^{7}$          & $-2.562 \times 10^{7}$          & $-2.564 \times 10^{7}$          & $-2.564 \times 10^{7}$          \\
\hline
\end{tabular}
  \caption{Table of log-likelihoods as $\nu$ and $s$ are
    varied. Minimums for each column are highlighted in \textbf{bold}.}
  \label{tab:nu-s-lml-table}
\end{table*}

\begin{figure*}[t]
  \centering
  \includegraphics[width=\textwidth]{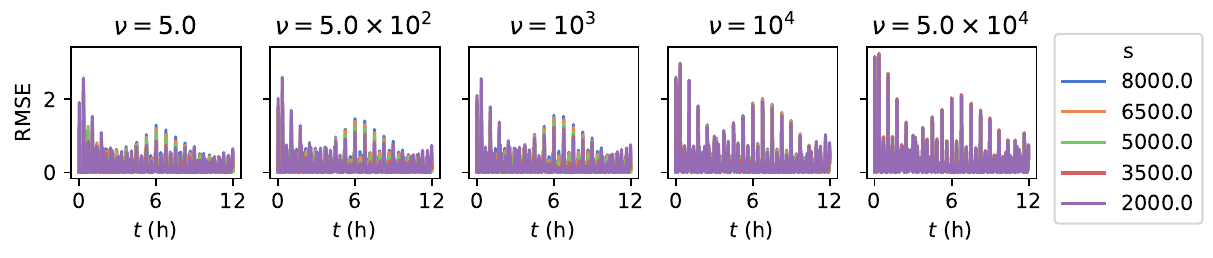}
  \caption{RMSE across time for the linear model, as $\nu$ and $s$ are varied.}
  \label{fig:linear-nu-s-rmse}
\end{figure*}

\begin{table*}[t]
  \centering
  \begin{tabular}{rlllll}
\hline
       & 2000.0                          & 3500.0                          & 5000.0                          & 6500.0                          & 8000.0                          \\
\hline
     5 & $\mathbf{-7.778 \times 10^{6}}$ & $\mathbf{-6.234 \times 10^{6}}$ & $\mathbf{-6.235 \times 10^{6}}$ & $\mathbf{-7.137 \times 10^{6}}$ & $\mathbf{-9.580 \times 10^{6}}$ \\
   500 & $-8.476 \times 10^{6}$          & $-6.901 \times 10^{6}$          & $-7.221 \times 10^{6}$          & $-8.630 \times 10^{6}$          & $-1.075 \times 10^{7}$          \\
  1000 & $-9.182 \times 10^{6}$          & $-7.653 \times 10^{6}$          & $-8.136 \times 10^{6}$          & $-9.662 \times 10^{6}$          & $-1.166 \times 10^{7}$          \\
 10000 & $-1.790 \times 10^{7}$          & $-1.734 \times 10^{7}$          & $-1.781 \times 10^{7}$          & $-1.856 \times 10^{7}$          & $-1.914 \times 10^{7}$          \\
 50000 & $-2.219 \times 10^{7}$          & $-2.337 \times 10^{7}$          & $-2.361 \times 10^{7}$          & $-2.360 \times 10^{7}$          & $-2.357 \times 10^{7}$          \\
\hline
\end{tabular}
  \caption{Linear misspecification: table of log-likelihoods, as $\nu$
    and $s$ are varied, for fixed spatiotemporal observation frequencies.}
  \label{tab:linear-nu-s-lml-table}
\end{table*}

\subsection{Investigating parametric misspecification}
\label{sec:parameter-misspecification}

Following these results, we now investigate joint parametric
misspecification of $s \in \{2000, 3500, 5000, 6500, 8000\}$ and
$\nu \in \{5, 500, 10^3, 10^4, 5 \times 10^4\}$.
We set $n_y = 1$, and $k = 30$ ($1$ spatial location observed every
$30$ \si{\second}). In Figure~\ref{fig:nu-s-rmse-nlml-layout} (top)
the \gls*{RMSE} is shown for the estimated posterior distributions
$p(\vu_m, \bm{\eta}_m \given \vy_{1:m}, \nu, s, \bm{\Theta}, \sigma)$.
As previous we see that the models with small $\nu$ are more
accurate. Additionally, as $s$ is increasingly misspecified there is a
stratification of model performance, with, unsurprisingly, the
correctly specified $s = 2000$ quite noticeably out-performing the
misspecified models. With larger $\nu$ values we see that there
is a mild increase in the \gls*{RMSE} through conditioning on
data. Less stratification appears to be present as $\nu$ is increased;
damping dominates the misspecified bathymetry in terms of mismatch.

For additional model comparison, we use the log-likelihood. Due to the
structure of this problem we can write this via factorisation
\begin{multline*}
  \log p(\vy_{1:M} \given \nu, s, \bm{\Theta}, \sigma) = \\
    \log p(\vy_1 \given \nu, s, \bm{\Theta}, \sigma) \\
    + \sum_{m = 2}^M \log p(\vy_{m} \given \vy_{1:m - 1}, \nu, s, \bm{\Theta}, \sigma).
\end{multline*}
This can be approximated when running the \gls*{LR-ExKF}, due to the
Gaussian approximation. The individual likelihoods are of the
form
\begin{multline*}
  p(\vy_{m} \given \vy_{1:m - 1}, \nu, s, \bm{\Theta}, \sigma) = \\
  = \NPDF(\mH \hat{\bm{\mu}}_m, (\mH \hat{\mL}_m) (\mH \hat{\mL}_m)^\top + \sigma^2 \mI),
\end{multline*}
where
$p(\vu_{m} \given \vy_{1:m - 1}, \nu, s, \bm{\Theta}, \sigma)
= \NPDF(\hat{\bm{\mu}}_m, \hat{\mL}_m \hat{\mL}_m^\top)$. This is a
strictly proper scoring rule with respect to Gaussian
measure~\citep{gneiting2007strictly}. Intuitively, this is an
uncertainty-weighted scoring rule that punishes models which are more
certain about inaccurate predictions of the data, at each observation
time.

The log-likelihoods are shown, across time, in
Figure~\ref{fig:nu-s-rmse-nlml-layout}. The models stratify across
$s$ more obviously for the well-specified models, with less
stratification as $\nu$ is increased. All models show a
gradual decrease in the log-likelihood values over time; conditioning
on data results in more accurate models. Note also that similar to the
\gls*{RMSE} values (see also
Figure~\ref{fig:rmse-post-nu-single-tidal}) we see that there is the
same quasi-periodic behaviour as the tidal forcing begins to approach
$0$, resulting in decreases in the likelihood.

For visual comparison, the average \gls*{RMSE} values and the
log-likelihoods are shown in
Figure~\ref{fig:nu-s-rmse-lml-table-color}. As previous, we see that
with $\nu \geq 10^4$ there is a clear increase in the \gls*{RMSE}
marking a qualitative change in the dynamics. Similar model
stratification is seen for the \gls*{RMSE} as is for the
log-likelihood; in these examples they perform similarly as model
comparison metrics. These log-likelihoods are tabulated in
Table~\ref{tab:nu-s-lml-table}. Models with $\nu = 5$ are preferred
across each bathymetry.

Following the computations of the log-likelihoods, we can perform
model comparison via Bayes factors~\citep{kass1995bayes}. The Bayes factor
is given by the ratio of the probabilities of the data given the
different assumed models:
\begin{multline}
  \label{eq:bayes-factors}
  \log \BF_{10}
  = \log p(\vy_{1:M} \given \nu_1, s_1, \bm{\Theta}, \sigma) \\
  - \log p(\vy_{1:M} \given \nu_0, s_0, \bm{\Theta}, \sigma).
\end{multline}
We see that there is strong evidence in favour of the well-specified
model in comparison to the others (smallest $\log \BF_{10} \approx 10^5$). In
each case it is clear that
increasing the degree of misspecification, by either shifting the topography,
or, increasing the misspecification, results in less performant models.
Models with smaller $\nu$ are preferred over those which have a larger
$\nu$. Interestingly, there is very strong evidence against the model
with $\{\nu = 5 \times 10^4, s=2000\}$, in comparison with that of
$\{\nu = 5, s = 8000\}$ ($\log \BF_{10} \approx
10^7$). We notice that the trend
misspecification due to $s$ tends to be less severe than that due to
$\nu$ (see also Figure~\ref{fig:nu-s-rmse-lml-table-color}).

\subsection{Linearisation}
\label{sec:linearisation}

Finally we investigate joint viscosity-bathymetry
misspecification as in the previous subsection, with the addition of
model linearisation. As previous, we vary
$s \in \{2000, 3500, 5000, 6500, 8000\}$ and
$\nu \in \{5, 500, 10^3, 10^4, 5 \times 10^4\}$, whilst fixing
$k = 30$ and $n_y = 1$ to compute the posterior estimates. We plot the
\gls*{RMSE} values across time for these linearised model
approximations in Figure~\ref{fig:linear-nu-s-rmse}. The \gls*{RMSE}
values, in comparison to those of the nonlinear models, are slightly
larger with notable increases in the cases of well-specified
bathymetry.

This disparity in model performance is further realised in the
log-likelihoods (seen in Table~\ref{tab:linear-nu-s-lml-table}) being
larger for the well-specified models in comparison to those of the
poorly specified models. We see the unsurprising results that
small-$\nu$ models perform better than the others. In comparing
Tables~\ref{tab:nu-s-lml-table}~and~\ref{tab:linear-nu-s-lml-table}
it is seen that when the severity of model misspecification is larger
(approximately $s \geq 5000$, $\nu \geq 10^4$) the linear model
outperforms the nonlinear model. For $s \geq 5000$ we posit this is
due to the ignoring of the resultant interactions between the
misspecified bathymetry and velocity. When damping is very highly
misspecified, even for a well-specified bathymetry the linear model is
preferred. Again this is thought to be due to the
addition of nonlinearity not really contributing to the dynamics ---
in this regime the dynamics are dominated by the linear dissipative
behaviour, in any case.

\section{Discussion and conclusion}
\label{sec:conclusion}

In this work we studied the efficacy of \gls*{statFEM} as applied to
the $1D$ \gls*{SWE}, to see how the methodology responds to scenarios
of increasing model misspecification. Previous work has necessarily
included smaller studies of milder cases of model misspecification;
this work provides the first systematic analysis of the approach under
gradually increased misspecification severity. Misspecification was
induced via linearisation, viscosity, and bottom-topography
(bathymetry), in regimes of reduced spatiotemporal observational
frequency. The \gls*{RMSE} and log-likelihood were used for model
comparison.

The methodology is able to appropriately deal with model
misspecification with notably large improvements in model error as the
number of observation locations is increased; the method performs well
in recovering misspecified dynamics. This is thought to be due to
spatial variation being more informative to the model error than
increasing the frequency of observations. The changes in observation
frequency are small in comparison to the timescale of the flow and
thus the differences in the observations, arriving at different times
are not large enough to warrant drastic reductions in model error
(though there is still a reduction). However, as wavelengths are
relatively long the additional information included via spatial
variation, through additional observation locations, does indeed
result in marked reductions in model error. Note also that our model
error term, the \gls*{GP} $\xi$, induces spatial correlations over
components of model error. Therefore including additional observation
locations, which make use of this error structure is again thought to
be helpful. We note that whilst we did not include temporal
correlation in $\xi$, \aprio{}, the additional study and comparison of
model error structures being correlated in both space and time is of
interest.

The effects of misspecification have different qualitative
behaviors. When $\nu$ is well-specified the bathymetry parameter $s$
results in immediate increases in error which are of similar
magnitude. On the other hand, as noted previously (see also
Figure~\ref{fig:rmse-post-nu-single-tidal}) when $\nu$ is misspecified
a regular quasi-periodic pattern in the error emerges which results
in nearly visually indistinguishable error patterns. We see, also,
that due to the domination of the dissipation these periodic-type
patterns in the error are also seen in the linear model for lower
values of $\nu$. More severe mismatch is seen to result from large
dissipation values rather than large shifts of the
bottom-topography. Both, however, are reduced through observing more
spatial locations. It is worth noting that there is a clear visual
decrease in the amount of model error present when taking $n_y = 2$
instead of $n_y = 1$ (see Figure~\ref{fig:nt-skip-nx-obs-rmse}), when
the topography is misspecified. When designing observation systems
(i.e. measurement/sensor locations) this suggests that taking
additional observation locations is valuable when they are of a
similar lengthscale to the flow under consideration. In cases of
severe misspecification linear approximations aid in slightly reducing
the model error, as seen via the log-likelihoods.

Whilst the \gls*{RMSE} and log-likelihood are useful and theoretically
sound metrics, the study of appropriate additional metrics (such as,
e.g., the Brier score~\citep{brier1950verification}) would be a useful
tool for practitioners when implementing and diagnosing models. We
also note that our results are conditioned on sets of \gls*{GP}
hyperparameters which, whilst chosen to ensure appropriate UQ on a
well-specified model, are not optimal with regards to the
log-likelihood. Joint investigation of hyperparameter estimation and
filtering is of interest and is a possible avenue of further
research.

Model error in this study appears to arrive in similar timescales no
matter which parameter is misspecified. In exploring alternate models
before we settled on the model used in this paper, we found that there
were intuitive interactions between the timescales of model error and
posterior updating. When mismatch occurs in fast timescales more
frequent updating is preferred. For slow timescales less frequent updating
is required.

These results provide additional evidence that the \gls*{statFEM}
approach allows for statistically coherent inference in regimes of
potentially severe model misspecification. The admission of spatially
correlated and physically sensible uncertainty results in improvements
in model accuracy as data is assimilated. The induced uncertainty is
sensible and reflects modelling choices: for example, boundary
conditions are respected and unobserved components are less certain
\apost{}. From the statistical point-of-view, the inclusion of
physical information alongside the \gls*{GP} enables the use of sparse
data. Results suggest that the inclusion of data, irrespective of the
amount, only aids in model proficiency when using \gls*{statFEM}.

\appendix

\section{FEM discretisation}
\label{sec:fem-convergence}

\begin{figure*}[t]
  \centering
  \includegraphics[width=0.5\textwidth]{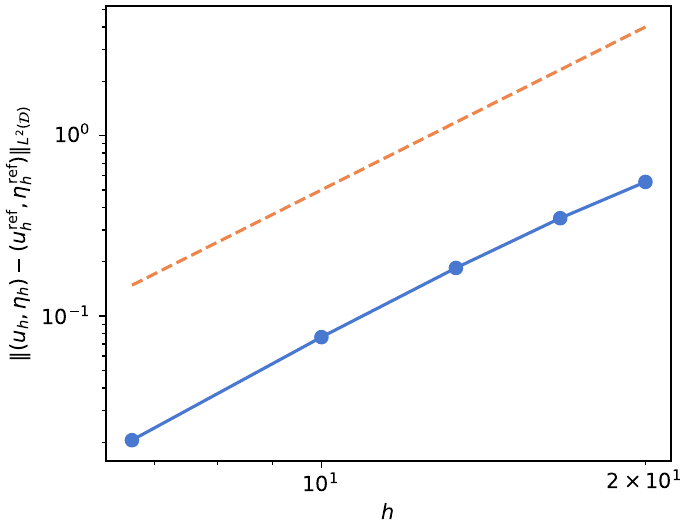}
  \caption{Deterministic \gls*{FEM} convergence for our employed
discretisation. Discretisation error is shown as a
    \textcolor{tab_blue}{blue line}~(\textcolor{tab_blue}{\rule[1.5pt]{0.4cm}{1pt}}),
    and dashed line shown in \textcolor{tab_orange}{orange}
    (\textcolor{tab_orange}{\rule[1.5pt]{0.25cm}{1pt}}~\textcolor{tab_orange}{\rule[1.5pt]{0.25cm}{1pt}})
  is for reference, having gradient 3.}
  \label{fig:fem-convergence}
\end{figure*}

To justify the chosen discretisation and level of mesh-refinement, the
deterministic \gls*{FEM} convergence results are plotted in
Figure~\ref{fig:fem-convergence}. We run a reference model
$(u_h^{\mathrm{ref}}, \eta_h^{\mathrm{ref}})$ with $n_v = 3000$
cells, and compute the $L^2$ errors against this reference
model, after running the models with $\Dt = 1$ up to time $t = 600$ \si{\second}
with meshes having $n_v \in \{ 500, 600, 750, 1000, 1500 \}$. Errors
shrink with a cubic rate (est. gradient $3.0144$).

\section{Notes on linear \acrshort*{statFEM}}
\label{sec:linear-model-derivation}
In the linear case, we recall that the \gls*{statFEM} model definition
is
\begin{equation}
  \begin{cases}
    u_t + g \eta_x + \nu u_{xx} = \xi_u, & x \in \cD, \\
    \eta_t + \left(H u \right)_{x} = \xi_\eta, & x \in \cD, \\
    u_x = 0, \; \eta = 0, & x \in \partial \cD.
  \end{cases}
\end{equation}
As previous we model the forcing terms $\xi_u$ and $\xi_\eta$ by
\aprio{} uncorrelated \glspl*{GP}. Making use of the same $P2$-$P1$ discretisation
as previous gives
\begin{align}
  \label{eq:swe-linear-discretised}
\mM_u \frac{\vu^{n} - \vu^{n - 1}}{\Dt} + \nu \mA \vu^{n - \theta} + g \mB \bm{\eta}^{n - \theta} &= \frac{1}{\sqrt{\Dt}}\bm{\xi}_u^{n - 1}, \nonumber \\
\mM_\eta \frac{\bm{\eta}^n - \bm{\eta}^{n - 1}}{\Dt} + \mB(H)\vu^{n - \theta} &= \frac{1}{\sqrt{\Dt}} \bm{\xi}_\eta^{n - 1},
\end{align}
where we have recycled the notation for the operators as in the main text.
Here we also have $\mB_{ji}(H) = \langle H \phi_{i, x} + H_x (H)
\phi_i, \psi_j \rangle$. The filtering procedure proceeds as previous,
where now instead of using a linearised approximation to the
prediction step, we compute this exactly (as now the Jacobian of the
r.h.s. of~\eqref{eq:swe-linear-discretised} does not depend on the
state $(\vu^n, \bm{\eta}^n)$). This is because we can write the linear
updating rule for the state as
\begin{align*}
  \mU_n \vu^n &= \mU_{n - 1} \vu^{n - 1} + \sqrt{\Dt} \bm{\xi}_u^{n - 1}, \\
  \mV_n \bm{\eta}^n &= \mV_{n - 1} \bm{\eta}^{n - 1} + \sqrt{\Dt} \bm{\xi}_\eta^{n - 1},
\end{align*}
where $\mU$, $\mV$ are defined as appropriately from \eqref{eq:swe-linear-discretised}.
For computation we employ the same low-rank approximation over the \glspl*{GP} $\xi_u$ and $\xi_\eta$.
Hyperparameters for these are the same as those used for the nonlinear model.
Inference in this scenario now proceeds via a standard low-rank \textit{Kalman filter}~\citep{kalman1960new} instead of the extended Kalman filter employed for the nonlinear models.

{\small \noindent\textbf{Data accessibility}: All code and data used in this work is publicly available on GitHub \texttt{\url{https://github.com/connor-duffin/sswe}}.}

{\small \noindent\textbf{Acknowledgements}: The authors would like to thank Bedartha Goswami and Youssef Marzouk for helpful discussions.}

{\small \noindent\textbf{Funding information}: C. Duffin and M. Girolami were supported by EPSRC grant EP/T000414/1.  E. Cripps, M. Girolami, M. Rayson and T. Stemler are supported by the ARC ITRH for Transforming energy Infrastructure through Digital Engineering (TIDE, \url{http://TIDE.edu.au}) which is led by The University of Western Australia, delivered with The University of Wollongong and several other Australian and International research partners, and funded by the Australian Research Council, INPEX Operations Australia, Shell Australia, Woodside Energy, Fugro Australia Marine, Wood Group Kenny Australia, RPS Group, Bureau Veritas and Lloyd's Register Global Technology (Grant No. IH200100009). M. G was supported by a Royal Academy of Engineering Research Chair, and EPSRC grants EP/W005816/1, EP/V056441/1, EP/V056522/1, EP/R018413/2, EP/R034710/1, and EP/R004889/1. E. Cripps was supported by Australian Research Council Industrial Transformation Training Centre (Grant No. IC190100031).}

{\small \noindent\textbf{Competing interests}: The authors have no competing interests to declare.}

{\small \noindent\textbf{Authors' contributions}: C. Duffin conceptualised the research, developed the code-base, ran the experiments, and wrote the manuscript. P. Branson, M. Rayson, E. Cripps, and T. Stemler conceptualised the research and revised the manuscript. M. Girolami conceptualised the research.}

\bibliographystyle{agsm}
\bibliography{bibliography.bib}

\end{multicols}

\end{document}